\documentclass[11pt,a4paper]{amsart}

\usepackage{times}
\usepackage{amsfonts,amsmath,amssymb}
\usepackage[mathcal]{euscript}

\usepackage{mathptmx}

\DeclareSymbolFont{SY}{U}{psy}{m}{n}
\DeclareMathSymbol{\emptyset}{\mathord}{SY}{'306}

\numberwithin{equation}{section}

\DeclareMathOperator*{\bigtimes}{\times}
\DeclareMathOperator{\diag}{diag}
\DeclareMathOperator{\Hom}{Hom}

\DeclareMathOperator{\Ran}{Ran}
\DeclareMathOperator{\Ker}{Ker}

\DeclareMathOperator{\sign}{sign}

\newcommand{\R}{\mathbb{R}}
\newcommand{\C}{\mathbb{C}}
\renewcommand{\SS}{\mathbb{S}}
\newcommand{\T}{\mathbb{T}}
\newcommand{\Z}{\mathbb{Z}}

\newcommand{\1}{\mathbb{I}}

\newcommand{\cA}{{\mathcal A}}

\newcommand{\cC}{{\mathcal C}}
\newcommand{\cD}{{\mathcal D}}
\newcommand{\cE}{{\mathcal E}}

\newcommand{\cG}{{\mathcal G}}
\newcommand{\cH}{{\mathcal H}}
\newcommand{\cI}{{\mathcal I}}

\newcommand{\cK}{{\mathcal K}}
\newcommand{\cL}{{\mathcal L}}
\newcommand{\cM}{{\mathcal M}}

\newcommand{\cU}{{\mathcal U}}

\newcommand{\U}{\mathsf{U}}

\newcommand{\W}{\mathsf{W}}

\newcommand{\vS}{$\mathrm{\Check{S}}$}

\newfont{\timitfont}{pplbi7t scaled 1000}
\newfont{\timitfontsmall}{pplbi7t scaled 700}
\renewcommand{\v}{\mbox{\timitfont{v}}}
\newcommand{\vsmall}{\mbox{\timitfontsmall{v}}}

\renewcommand{\i}{\mathrm{i}}
\newcommand{\e}{\mathrm{e}}

\newtheorem{theorem}{Theorem}[section]%
\newtheorem{lemma}[theorem]{Lemma}%
\newtheorem{proposition}[theorem]{Proposition}%
\newtheorem{remark}[theorem]{Remark}%
\newtheorem{definition}[theorem]{Definition}
\newtheorem{example}[theorem]{Example}{\bf}{\rm}

\begin{document}

\newtheorem{introtheorem}{Theorem}{\bfseries}{\itshape}%
\newtheorem{introcorollary}[introtheorem]{Corollary}{\bfseries}{\itshape}%
\newtheorem{introdefinition}[introtheorem]{Definition}{\bfseries}{\itshape}%

\title{Quantum Wires with Magnetic Fluxes}

\author[V.~Kostrykin]{Vadim Kostrykin}
\address{Vadim Kostrykin\\ Fraunhofer-Institut f\"{u}r Lasertechnik, Steinbachstra{\ss}e
15, D-52074\\ Aachen, Germany} \email{kostrykin@ilt.fraunhofer.de,
kostrykin@t-online.de}

\author[R.~Schrader]{Robert Schrader$^\ast$}
\address{Robert Schrader\\ Institut f\"{u}r Theoretische Physik\\
Freie Universit\"{a}t Berlin, Arnimallee 14\\ D-14195 Berlin, Germany}
\email{schrader@physik.fu-berlin.de}
\thanks{$^\ast$ Supported in part by DFG SFB
288 ``Differentialgeometrie und Quantenphysik''}

\date{29 November 2002}

\dedicatory{Dedicated to Rudolf Haag in honor of his 80th birthday}


\renewcommand{\thefootnote}{\fnsymbol{footnote}}
\footnotetext{Published in Comm. Math. Phys. (2003)}


\maketitle

\begin{abstract}
In the present article magnetic Laplacians on a graph are analyzed. We
provide a complete description of the set of all operators which can be
obtained from a given self-adjoint Laplacian by perturbing it by magnetic
fields. In particular, it is shown that generically this set is isomorphic
to a torus. We also describe the conditions under which the operator is
unambiguously (up to unitary equivalence) defined by prescribing the
magnetic fluxes through all loops of the graph.
\end{abstract}

\section{Introduction and Main Results}

Magnetic Laplacians on finite graphs appear in a number of physical
applications. The major interest in operators of this type originates from
the study of quantum transport in mesoscopic networks (see, e.g.,
\cite{ACDMT}, \cite{Avron:Raveh:Zur}, \cite{Avron:Raveh:Zur:2},
\cite{Avron:Sadun}, \cite{AS}, \cite{AEGS}, \cite{Buettiker},
\cite{Chalker-Coddington}). Magnetic Laplacians have also been the subject
of several studies in the context of quantum chaos
\cite{Kottos_Smilansky1}, \cite{Kottos_Smilansky2},
\cite{Kottos_Smilansky3}. Graph-theoretical generalizations of the Harper
operator provide discrete models describing the dynamics of a quantum
particle in the presence of a magnetic field (see \cite{Sunada},
\cite{Mathai} and references therein).

The most intriguing feature of Laplacians on graphs is the relation between
their spectral properties and the geometry of the graph. For
\emph{discrete} Laplacians the study of this relation is one of the central
issues of spectral graph theory (see, e.g., \cite{Chung}). As an example,
the multiplicity of the zeroth eigenvalue of the discrete Laplacian equals
the number of connected components of the graph. In the presence of an
external magnetic field the effect of the topology of the graph becomes more
pronounced. Thus, the quantum conductance in networks is known to be related
to Chern numbers \cite{Avron:Raveh:Zur}, \cite{Avron:Raveh:Zur:2},
\cite{Avron:Sadun}, \cite{AS}.

In the present article we study general \emph{differential} self-adjoint
magnetic Laplacians on finite graphs. This work is a continuation of
our previous studies \cite{KS}, \cite{KS2}, \cite{KS3}.

We consider an arbitrary nontrivial connected (metrical) graph $\cG$ with a finite number $n\geq
0$ of external lines and a finite number $m\geq 0$ of internal lines (henceforth also called
edges), $m+n\neq 0$. More precisely, this means that outside of a finite domain the graph is
isomorphic to the union of $n$ positive half-lines. Any internal line ends at two, not necessary
different vertices and has a finite length. We assume that any vertex $\v$ of $\cG$ has non-zero
degree $\deg(\v)$, i.e., for any vertex there is at least one edge (internal or external) with
which it is incident.

Let the set $\cE\;(|\cE|=n)$ label the external and the set
$\cI\;(|\cI|=m)$ the internal lines of the graph. We assume that the sets
$\cE$ and $\cI$ are ordered in an arbitrary but fixed way. To each
$e\in\cE$ we associate the semi-infinite interval $[0,\infty)$ and to each
$i\in\cI$ the finite directed interval $[0,a_i]$, where $a_i>0$ is the
length of this line. With this association the graph becomes directed, such
that the initial vertex of an edge of length $a_{i}$ corresponds to $x=0$
and the terminal vertex corresponds to $x=a_{i}$. The external lines are
assumed to be directed in the positive direction of half-lines. We denote
by $\underline{a}$ the collection of the lengths $\{a_i\}_{i\in\cI}$.

Though not necessary but rather as a motivation for our approach we
temporarily assume that the graph is isometrically imbedded in three
dimensional Euclidean space $\R^3$. Given a vector potential
$\mathbf{A}\in\left(C(\R^3)\right)^3$, whose curl gives a magnetic field,
the question arises what is the reasonable self-adjoint operator describing
the evolution of quantum states on the graph in the presence of this
external magnetic field. A common intuitive construction goes as follows. We
assume that all external lines and all edges of the imbedded graph $\cG$
are oriented smooth curves. Let $\tau_j(x)\;(j\in\cE\cup\cI)$ denote the
unit tangential field on the $j$-th edge of the graph chosen in accordance
with the orientation of the edge. We set $\cA_j (x) = \langle\tau_j(x),
\mathbf{A}(x)\rangle$ where $\langle\cdot,\cdot\rangle$ denotes the inner
product in $\R^3$. The vector potential leads to the differential
expressions $\left(-\i\frac{d}{dx}-\cA_j(x)\right)^2$ on the external lines
or edges of the graph with appropriate boundary conditions at the vertices.
More precisely, consider the family $\psi=\{\psi_{j}\}_{j\in\cE\cup\cI}$ of
complex valued functions defined on $[0,\infty)$ if $j\in\cE$ and on
$[0,a_{i})$ if $j\in\cI$. Formally the Laplace opertor is defined as
\begin{equation}\label{diff:expression}
\left(\Delta(A,B;\cA)\psi\right)_j (x) = \left(\frac{d}{dx} - \i
\cA_j(x)\right)^2 \psi_j(x),\qquad j\in\cI\cup\cE
\end{equation}
with the boundary conditions
\begin{equation}
\label{Randbedingungen}
A\underline{\psi} + B(\underline{\psi}'-\i\underline{\cA}\underline{\psi}) =
0.
\end{equation}
Here
\begin{equation*}
\underline{\psi} = \begin{pmatrix} \{\psi_e(0)\}_{e\in\cE} \\
                                   \{\psi_i(0)\}_{i\in\cI} \\
                                   \{\psi_i(a_i)\}_{i\in\cI} \\
                                     \end{pmatrix},\qquad
\underline{\psi}' = \begin{pmatrix} \{\psi_e'(0)\}_{e\in\cE} \\
                                   \{\psi_i'(0)\}_{i\in\cI} \\
                                   \{-\psi_i'(a_i)\}_{i\in\cI} \\
                                     \end{pmatrix},
\end{equation*}
and $\underline{\cA}$ is the diagonal matrix
\begin{equation}
\label{a:def} \underline{\cA} = \diag(\{\cA_{e}(0)\}_{e\in\cE},\{\cA_{i}(0)\}_{i\in\cI},
\{-\cA_{i}(a_{i})\}_{i\in\cI})
\end{equation}
with the same ordering as used for $\underline{\psi}$ and $\underline{\psi}'$.
If $\cA=0$ we will simply write $\Delta(A,B)$
instead of $\Delta(A,B;0)$.

Actually we do not need to imbed the graph in $\R^3$ and we may simply
prescribe the vector potentials $\cA(x)=\{\cA_j(x)\}_{j\in\cI\cup\cE}$ on
all external lines and edges of the graph. So having given this motivation,
our further discussion will be completely intrinsic.

Below we will prove (Theorem \ref{thm:3.1}) that the operator
$\Delta(A,B;\cA)$ is self-adjoint if and only if $A B^\dagger$ is symmetric
(with $\dagger$ denoting the Hermitian conjugation) and the
$(n+2m)\times(n+2m)$ matrix $(A,B)$ has a maximal rank. This is an
extension of a result in \cite{KS} to the case of magnetic Laplacians.

The \textit{local} gauge transformation
\begin{equation}\label{gauge:trans}
G:\quad\psi_j(x) \mapsto \psi_j(x) \exp\left\{\i \chi_j(x)\right\},\qquad
j\in\cI\cup\cE
\end{equation}
with
\begin{equation}\label{intro:diff}
\chi'_j(x)=\cA_j(x)
\end{equation}
eliminates the vector potential in the differential expression
\eqref{diff:expression} and in the boundary conditions
\eqref{Randbedingungen}. However, the gauge transformation changes the
boundary condition: It transforms the magnetic
Laplacian $\Delta(A,B;\cA)$ to the Laplacian $\Delta(A\cU,B\cU)$ without
magnetic field,
\begin{equation*}
G^{-1}\Delta(A,B;\cA)G = \Delta(A\cU,B\cU),
\end{equation*}
where the transformation $\cU\equiv\cU_G$ is given by the diagonal, unitary
matrix acting on the boundary values $\underline{\psi}$ and
$\underline{\psi}'$
\begin{equation}\label{intro:u}
\cU = \diag\left(\left\{\e^{\i \chi_e(0)}\right\}_{e\in\cE}, \left\{\e^{\i
\chi_i(0)}\right\}_{i\in\cI},\left\{\e^{\i \chi_i(a_{i})}\right\}_{i\in\cI}\right)
\end{equation}
with
\begin{equation*}
\chi_i(a_i) := \chi_i(0) + \int_0^{a_i} \cA_i(t) dt,\qquad i\in\cI.
\end{equation*}

In the other words, the perturbation of the Laplacian $\Delta(A,B)$ by the
vector potential $\cA$ is equivalent to the transformation of the boundary
conditions $(A,B) \mapsto (A\cU,B\cU)$ by some diagonal unitary matrix
$\cU$. Conversely, given a diagonal unitary $(n+2m)\times(n+2m)$ matrix
$\cU$ there is a vector potential $\cA$ (possibly zero) such that
$\Delta(A\cU,B\cU)$ can be obtained from $\Delta(A,B;\cA)$ by means of a
local gauge transformation. We will call $\Delta(A\cU,B\cU)$ a {\it
magnetic perturbation} of $\Delta(A,B)$.

This correspondence can be described more precisely in a group-theoretical
setting. Let $\U=\U(\cG)$ denote the group of all diagonal unitary
$(n+2m)\times (n+2m)$ matrices enumerated by the elements of the sets $\cE$
and $\cI$ in the same ordering as in \eqref{intro:u}. Let $\U_0$ be its
subgroup consisting of those matrices with the structure \eqref{intro:u}
with $\exp\{\i \chi_i(0)\}=\exp\{\i \chi_i(a_i)\}$ for any $i\in\cI$.
Equivalently, the elements in $\U_0$ correspond to gauge transformations
\eqref{gauge:trans} with vanishing vector potential $\cA$. With this
notation the cosets $\U/\U_0$ of $\U_0$ in $\U$ are, obviously, in a
one-to-one correspondence with the points $\{t_{i}\}_{i\in\cI}$ of the
$m$-dimensional torus $\T^m$ given as
\begin{equation*}
t_i=\exp\left\{\i \int_0^{a_i} \cA_i(x) dx\right\}\in\SS, \qquad i\in\cI.
\end{equation*}

The first main result of the present work provides an answer to the
following question
\begin{itemize}
\item[$\bullet$] What is the orbit $O(A,B):=\{\Delta(A\cU,B\cU)|\ \cU\in\U \}$ of the operator
$\Delta(A,B)$ in the set of all self-adjoint Laplacians under the action of
the group $\U$? Roughly speaking, the question is how many
\emph{different} magnetic perturbations of a given Laplacian $\Delta(A,B)$
there are.
\end{itemize}

The orbit of $\Delta(A,B)$ under the action of $\U$ is isomorphic to the factor group
$\U/\W(A,B)$, where $\W(A,B)$ is the isotropy group (or stabilizer) of the point $\Delta(A,B)$,
i.e. the subgroup which leaves $\Delta(A,B)$ invariant. Recall that by a result of \cite{KS}
$\Delta(A\cU,B\cU)=\Delta(A,B)$ if and only if there is an invertible $(n+2m)\times(n+2m)$ matrix
$C$ such that $CA=A\cU$ and $CB=B\cU$.

The following theorem provides an explicit description of $\W(A,B)$. Let $A^\star$ denote the
Moore-Penrose pseudoinverse of $A$ (see, e.g., \cite{KS} for a short presentation of the main
facts related to this notion).

\begin{introtheorem}\label{x:theorem:intro:1}
The element $\cU\in\U$ belongs to $\W(A,B)$ if and only if
\begin{itemize}
\item[(i)] $\cU$ leaves both $\Ker A$ and $\Ker B$ invariant;
\end{itemize}
and
\begin{itemize}
\item[(ii)] one of the following equivalent commutator relations
\begin{equation}\label{x:compat:intro:1}
[\cU, A^\star B]z = 0 \quad\text{for any}\quad z\in B^{-1}(\Ran A \cap \Ran
B)
\end{equation}
or
\begin{equation}\label{x:compat:intro:2}
[\cU, B^\star A]z = 0 \quad\text{for any}\quad z\in A^{-1}(\Ran A \cap \Ran
B)
\end{equation}
is fulfilled.
\end{itemize}

In particular, if $\Ran A \cap \Ran B =\{0\}$ then
$\Delta(A\cU,B\cU)=\Delta(A,B)$ for all $\cU\in \U$ such that $\cU$ leaves
both $\Ker A$ and $\Ker B$ invariant.
\end{introtheorem}

Note that since $\cU$ is unitary, the condition (i) implies that both
orthogonal decompositions $\Ker A\oplus[\Ker A]^\perp$ and $\Ker
B\oplus[\Ker B]^\perp$ of $\C^{n+2m}$ reduce $\cU$. Also we remark that
$\Ker A\cap \Ker B=\{0\}$ since $(A,B)$ has maximal rank.

There are examples of boundary conditions $(A,B)$ (see, e.g., Example \ref{Dirichlet} below) where
$\W(A,B)=\U$ and, thus, the orbit $O(A,B)$ consists of the only point $\Delta(A,B)$.

An important class of boundary conditions are \emph{local} boundary
conditions. They are such that they
couple only those boundary values of $\psi$ and of its derivative $\psi'$
which belong to the same vertex. The precise definition is as follows.

Let $S(\v)\subseteq \cE\cup\cI$ denote the star graph of the vertex $\v\in
V$, i.e., the set of the edges adjacent to $\v$. Also, by $S_{-}(\v)$
(respectively $S_{+}(\v)$) we denote the set of the edges for which $\v$ is
the initial vertex (respectively terminal vertex). Obviously, $S_{+}(\v)\cap
S_{-}(\v)=\emptyset$ if $\cG$ does not contain a cycle of length 1, i.e. a
tadpole.

Assume the elements $z$ of the linear space $\C^{n+2m}$ are written as
\begin{equation}\label{elements}
z=(\{z_e\}_{e\in\cE}, \{z^{(-)}_i\}_{i\in\cI}, \{z^{(+)}_i\}_{i\in\cI})^T.
\end{equation}
Consider the orthogonal decomposition
\begin{equation*}
\C^{n+2m} = \bigoplus_{\v\in V} \cL_{\v}
\end{equation*}
with $\cL_{\v}$ being the linear subspace of dimension $\deg(\v)$ spanned
by those elements \eqref{elements} of $\C^{n+2m}$ which satisfy
\begin{equation}
\label{decomp}
\begin{split}
z_e=0 &\quad \text{for all}\quad e\in \cE\setminus S(\v),\\
z^{(-)}_i=0 &\quad \text{for all}\quad i\in \cI\setminus S_-(\v),\\
z^{(+)}_i=0 &\quad \text{for all}\quad i\in \cI\setminus S_+(\v).
\end{split}
\end{equation}

\begin{introdefinition}\label{def:local}
Given the graph $\cG$, the boundary conditions $(A,B)$ are called
\emph{local} if there is an invertible $(n+2m)\times(n+2m)$ matrix $C$ and
linear transformations $A_{\v}$ and $B_{\v}$ in $\cL_{\v}$ such that the
simultaneous direct sum decompositions
\begin{equation}\label{permut}
CA= \bigoplus_{\v\in V} A_{\v}\quad \text{and}\quad CB= \bigoplus_{\v\in V} B_{\v}
\end{equation}
are valid.

Otherwise the boundary conditions are \emph{non-local}.
\end{introdefinition}

As outlined in \cite{KS} (see also \cite{KuSe}) for an arbitrary
self-adjoint boundary condition $(A,B)$ there is always a graph with
external lines labeled by $\cE$, internal lines labeled by $\cI$ with a set
of lengths $\underline{a}$, for which these boundary conditions are local.
This graph is unique under the requirement that the number of vertices be
maximal. We will elaborate on this in \cite{KS4}.

Consider the group $\W_0$ consisting of those elements $\cU$ of $\U$ which
admit a decomposition
\begin{equation*}
\cU = \bigoplus_{\v\in V} \cU_{\v},\qquad
\cU_{\v}:\,\cL_{\v}\rightarrow\cL_{\v}
\end{equation*}
with $\cU_{\v}$ being a multiple of the $\deg(\v)\times\deg(\v)$ unit matrix (see Definition
\ref{W0} below). It is an immediate consequence of Theorem
\ref{x:theorem:intro:1} that $\W_0$ is a subgroup of the isotropy group $\W(A,B)$ for any local
(self-adjoint) boundary conditions. For non-local boundary conditions it may well happen that
$\W(A,B)\subsetneq \W_0$ (see Example \ref{ex:4.1} below).

The next question we turn to  is
\begin{itemize}
\item[$\bullet$] What is the correspondence between the points in the orbit of the operator
$\Delta(A,B)$ under the action of the group $\U$ and the magnetic fluxes through the loops (i.e.,
closed paths) of the graph $\cG$? Given a Laplacian $\Delta(A,B)$ can its magnetic perturbation
$\Delta(A\cU,B\cU)$ be unambiguously defined by prescribing the magnetic fluxes through all loops
of the graph?
\end{itemize}

We will give an affirmative answer in Theorem \ref{intro:thm:2} below. To formulate this result we
need some additional notation.

A vector potential $\cA=\{\cA_j\}_{j\in\cE\cup\cI}$ defines a \emph{flux
map}, by which we associate to any loop $\gamma$ in the graph $\cG$ the
complex exponential of the magnetic flux through $\gamma$,
\begin{equation}\label{intro:flux}
\Phi_{\cA}(\gamma) = \exp \left\{\i \Big(\sum_i \sign(i) \int_0^{a_i}
\cA_i(x) dx\Big)\right\}.
\end{equation}
Here the sum is taken over all edges $i\in\cI$ constituting the loop
$\gamma$ with $\sign(i)=+1$ if the orientation of $\gamma$ coincides with
that of the edge $i$ and $\sign(i)=-1$ otherwise. If the graph $\cG$ is
imbedded in $\R^3$, by Stokes theorem the expression in the brackets
$(\ldots)$ has indeed the meaning of a flux through any oriented bordered
manifold $M$ with boundary $\partial M = \gamma$.

By established equivalence between vector potentials and transformation matrices $\cU\in\U$ the
flux map \eqref{intro:flux} can be alternatively expressed in terms of the matrix elements of
$\cU$ \eqref{intro:u} as
\begin{equation}\label{intro:flux:prime}
\Phi_{\cU}(\gamma) = \exp \left\{ \i \Big(\sum_i \sign(i)
(\chi_i(a_i)-\chi_i(0))\Big)\right\}
\end{equation}
and where the sum is as in \eqref{intro:flux}.

Loops of the graph $\cG$ can be realized as elements of the additive Abelian
group $H_1(\cG,\Z)$, the first homology group of the graph $\cG$. Recall
that this group is the Abelianization of the fundamental group $\pi_1(\cG)$.
Its elements are formal linear combinations with integer coefficients of the
``basis'' loops in the graph,
\begin{equation*}
c=\sum_{p} n_p c_p,\qquad n_p\in\Z.
\end{equation*}

For any $\cU\in\U$ the flux map $\Phi_{\cU}$ defined by
\eqref{intro:flux:prime} can uniquely be extended to a group
homomorphism from $H_1(\cG,\Z)$ to
$\SS$. The set of all such homomorphisms forms an Abelian group
denoted by $\Hom(H_1(\cG,\Z), \SS)$. The map
$\rho:\ \cU\mapsto \Phi_{\cU}$ is obviously a group homomorphism. Actually,
it is an epimorphism (Lemma \ref{epimorphism}). It is clear that
$\U_0\subseteq\Ker\,\rho$. In Lemma \ref{thm:4.3} below we will prove that
$\Ker\,\rho=\U_0 \W_0$.

The second main result of the present work is given by the following
theorem.

\begin{introtheorem}\label{intro:thm:2}
Assume that the self-adjoint boundary conditions $(A,B)$ are local. Then the groups $\U/(\U_0
\W(A,B))$ and $\Hom(H_1(\cG,\Z),\SS)/\rho(\W(A,B))$ are isomorphic. In particular, if
$\W(A,B)=\W_0$, then the groups $\U/(\U_0 \W_0)$ and $\Hom(H_1(\cG,\Z),\SS)$ are isomorphic.
\end{introtheorem}

Whenever the local boundary conditions $(A,B)$ are such that
$\W(A,B)=\W_0$, Theorem \ref{intro:thm:2} states that for any given
magnetic flux there is a unique set (up to transformations defined by
elements of $\U_0$ and $\W_0$) of phase factors $\cU\in\U$ giving this
flux. The situation changes drastically for boundary conditions such that
$\W(A,B)$ is strictly larger than $\W_0$. The transformations $(A,B)\mapsto
(A\cU, B\cU)$ with elements $\cU$ of $\W$ which are not elements of $\W_0$
do not change the operator $\Delta(A,B)$. However, they change the magnetic
flux! In the latter case, however, there is still a subgroup of
$\Hom(H_1(\cG,\Z),\SS)$ corresponding to those fluxes which remain
unaffected by all transformations $(A,B)\mapsto (A\cU, B\cU)$ induced by
the group $\W$. This subgroup may be realized as a factor group
$\Hom(H_1(\cG,\Z),\SS)/\rho(\W)$. Example \ref{Dirichlet} below shows that
$\W$ may be as large as $\U$ such that $\Hom(H_1(\cG,\Z),\SS)/\rho(\W)$ is
trivial in this case.

According to Theorem \ref{intro:thm:2} magnetic Laplacians depend on the
magnetic flux through any loop of the graph $2\pi$-periodically. In the
special case of a ring (in our context a graph formed by a single internal
line with coinciding initial and terminal vertices) this fact is well known
in the physics literature as the Byers-Yang-Bloch Theorem
\cite{Byers:Yang}, \cite{Bloch}.

The following corollary of Theorem \ref{intro:thm:2} states that if $\cG$ is a tree then an
arbitrary magnetic field $\cA(x)$ does not change the Laplacian $\Delta(A,B)$ for any local
boundary conditions $(A,B)$ in the sense that there is a local gauge transformation $G$ such that
$\Delta(A\cU_G,B\cU_G)=\Delta(A,B)$ with $\cU_G$ being given by \eqref{intro:u}. The precise
statement is as follows.

\begin{introcorollary}\label{coro:1}
If $\pi_1(\cG)$ is trivial (and hence $H_1(\cG;\Z)$ is also trivial), i.e. if the graph $\cG$
contains no loops, then $\U=\U_0\W(A,B)$ for arbitrary local boundary conditions $(A,B)$.
\end{introcorollary}

The requirement of locality is crucial since non-local boundary conditions
may have the same effect as a loop. An example of such situation is
presented in Example \ref{ex:4.1} below.

As already mentioned magnetic Laplacians on graphs appear as models of
different physical systems. In particular, the Chalker-Coddington network
model [15] was designed to describe the semiclassical motion of a single
electron in the presence of a perpendicular uniform magnetic field and a
random potential. Thus, in this model the internal lines of the network
(graph) are determined by the equipotential lines of the potential and the
vertices by its saddle points where two equipotential lines closely
approach one another. Correspondingly, the tunneling and transmission
probabilities then give rise to boundary conditions at the vertices. As a
consequence of the results of the present work a full quantum version of
the Chalker-Coddington model can be developed.

\subsection*{Acknowledgement} One of the authors (R.S.) would like to thank
M.~Schmidt and E.~Vogt for advise and help. Discussions with
H.~Schulz-Baldes have been very helpful.

\section{Laplacians With Magnetic Field}\label{sec:2}

To the triple $(\cE,\cI,\underline{a})$ with $\cE$, $\cI$, and
$\underline{a}$ as being given in the Introduction we associate the Hilbert
space $\cH=\cH(\cE,\cI,\underline{a})$ given as the orthogonal sum
\begin{equation*}
\cH=\cH_{\cE}\ \oplus\ \cH_{\cI}, \qquad \cH_{\cE}=\bigoplus_{e\in\cE}\cH_{e},
\qquad \cH_{\cI}=\bigoplus_{i\in\cI}\cH_{i},
\end{equation*}
where $\cH_e=L^2(0,\infty)$ and $\cH_i=L^2(0,a_i)$. The inner product in $\cH$ is given by
\begin{equation*}
\langle \phi,\psi\rangle_{\cH} = \sum_{e\in\cE} \int_0^\infty
\overline{\phi_e(x)} \psi_e(x)\ dx
+ \sum_{i\in\cI} \int_0^{a_i}\overline{\phi_i(x)} \psi_i(x)\ dx.
\end{equation*}
Elements of $\cH$ are written as column vectors
\begin{equation}\label{graph:ext:int}
\psi=\begin{pmatrix}\{\psi_e\}_{e\in\cE}\\
\{\psi_i\}_{i\in\cI}\end{pmatrix}, \qquad \psi_e\in\cH_e,\qquad
\psi_i\in\cH_i.
\end{equation}
Similarly we define the Sobolev space
$W^{2,2}=W^{2,2}(\cE,\cI,\underline{a})$ as
\begin{equation*}
W^{2,2}=\bigoplus_{e\in\cE}W^{2,2}(0,\infty)\ \oplus\ \bigoplus_{i\in\cI}
W^{2,2}(0,a_i),
\end{equation*}
where $W^{2,2}(0,\infty)$ and $W^{2,2}(0, a_i)$ are the usual Sobolev
spaces of square integrable functions whose distributional second
derivatives are also square integrable (see, e.g., \cite{RS2}). We observe
that the Hilbert space is independent of the particular graph constructed
out of the set of data $(\cE,\cI,\underline{a})$.

\begin{theorem}\label{thm:3.1}
Let $\cA(x)=\{\cA_j(x)\}_{j\in\cI\cup\cE}$ be continuous and bounded. Any
two $(n+2m)\times(n+2m)$ complex matrices $A$ and $B$ satisfying
\begin{equation}\label{mag:sa}
A B^\dagger - B A^\dagger = 0
\end{equation}
and such that the $(n+2m)\times 2(n+2m)$ matrix $(A,B)$ has maximal rank
equal to $n+2m$, define a self-adjoint magnetic Laplacian
\begin{equation*}
\left(\Delta(A,B; \cA)\psi\right)_j (x) =
\left(\frac{d}{dx} - \i \cA_j(x)\right)^2 \psi_j(x),\qquad j\in\cI\cup\cE.
\end{equation*}
in $\cH$ corresponding to the boundary condition
\begin{equation}\label{2:magnetic}
A\underline{\psi}+B(\underline{\psi}^\prime -
\i\underline{\cA}\underline{\psi})=0
\end{equation}
and with the diagonal $(n+2m)\times(n+2m)$ matrix $\underline{\cA}$ given by \eqref{a:def}.

Conversely, any self-adjoint magnetic Laplacian
corresponds to the boundary condition \eqref{2:magnetic} with some matrices
$A$ and $B$ satisfying \eqref{mag:sa}.
\end{theorem}

\begin{proof}
Consider the symmetric operator $\Delta_\cA^0$ defined by
\begin{equation*}
\left(\Delta_\cA^0 \psi\right)_j (x) = \left(\frac{d}{dx} - \i\cA_j(x)\right)^2 \psi_j(x)
\end{equation*}
with domain $\cD(\Delta_\cA^0)\subset W^{2,2}$ consisting of functions
which vanish at the vertices together with their first derivative. It is
clear that $\Delta_\cA^0$ has defect indices $(k,k)$ with $k=|\cE|+2|\cI|$.

On $W^{2,2}$ we consider the Hermitian symplectic form
\begin{equation*}
\Omega_\cA (\phi,\psi) = \langle \Delta_\cA \phi, \psi\rangle - \langle \phi, \Delta_\cA
\psi\rangle = - \overline{\Omega_\cA (\psi,\phi)}
\end{equation*}
with $\Delta_\cA$ being considered as a formal differential expression.

Let $\left[\ \right]_{\cA}: W^{2,2}\rightarrow \C^{2(n+2m)}$ be the
surjective linear map which associates to each $\psi$ the element
$[\psi]_{\cA}$ given as
\begin{equation*}
[\psi]_{\cA} = \begin{pmatrix} \{\psi_e(0)\}_{e\in\cE}\\ \{\psi_i(0)\}_{i\in\cI}\\
\{\psi_i(a_i)\}_{i\in\cI}\\
\{\psi^\prime_e(0)- \i \cA_e(0) \psi_e(0)\}_{e\in\cE}\\ \{\psi^\prime_i(0)-
\i \cA_i(0)
\psi_i(0)\}_{i\in\cI}\\
\{-\psi^\prime_i(a_i)+ \i \cA_i(a_i) \psi_i(a_i)\}_{i\in\cI}
\end{pmatrix}=\begin{pmatrix}\underline{\psi} \\ \underline{\psi}^\prime-\i\underline{\cA}
\underline{\psi}
\end{pmatrix}.
\end{equation*}
If $\cA(x)$ is continuously differentiable then by means of partial integration one verifies
\begin{equation}\label{quadratic:form}
\begin{split}
\Omega_\cA (\phi,\psi) = & \sum_{e\in\cE} \left[\overline{\phi_e(0)} \psi'_e(0) -
\overline{\phi'_e(0)} \psi_e(0) -2\i\cA_e(0) \overline{\phi_e(0)} \psi_e(0)\right]\\
& + \sum_{i\in\cI} \left[\overline{\phi_i(0)} \psi'_i(0) -
\overline{\phi'_i(0)} \psi_i(0) -2\i\cA_i(0) \overline{\phi_i(0)} \psi_i(0)\right]\\
& - \sum_{i\in\cI} \left[\overline{\phi_i(a_i)} \psi'_i(a_i) -
\overline{\phi'_i(a_i)} \psi_i(a_i) -2\i\cA_i(a_i) \overline{\phi_i(a_i)} \psi_i(a_i)\right]\\
= & \langle [\phi]_{\cA}, J [\psi]_{\cA}\rangle_{\C^{2(n+2m)}},
\end{split}
\end{equation}
where
\begin{equation*}
J=\begin{pmatrix} 0 & \1 \\ -\1 & 0 \end{pmatrix}
\end{equation*}
is the canonical symplectic matrix on $\C^{2(n+2m)}$. Here $\1$ is the $(n+2m)\times(n+2m)$
unit matrix. In the general case we again obtain
\eqref{quadratic:form} if we approximate $\cA(x)$ by continuously
differentiable functions.

Let the linear subspace $\cM(A,B)$ of $\C^{2(n+2m)}$ be given as the set of all
\begin{equation*}
[\psi]_{\cA}= \begin{pmatrix} \underline{\psi} \\
\underline{\psi}'-\i\underline{\cA}\underline{\psi}\end{pmatrix} \in \C^{2(n+2m)}
\end{equation*}
satisfying
\begin{equation*}
A \underline{\psi} + B (\underline{\psi}'-\i\underline{\cA}\underline{\psi}) = 0.
\end{equation*}
All self-adjoint extensions of $\Delta^0_{\cA}$ are described by maximal
isotropic subspaces $\cM(A,B)$ \cite{KS}. By Lemma 2.2 of \cite{KS} we
obtain the claim of the theorem.
\end{proof}

\begin{remark}
The technique of using Hermitian symplectic forms in extension theory seems
to have been well known for a long time. The earliest reference we are
aware of is \cite[Section 10]{HP}. In the context of differential and
difference operators on graphs and in similar contexts this technique has
also been used in \cite{Pavlov}, \cite{Novikov}, \cite{H}, \cite{BG},
\cite{Pavlov:2}.

We also mention the articles \cite{Ca2} and \cite{Ca1} where the self-adjointness conditions
\eqref{mag:sa} were proven (without using the formalism of Hermitian symplectic forms) for the
cases $m=0$, $n$ arbitrary and $m=1$, $n=0$, respectively.
\end{remark}

The assumption that the vector potential is bounded formally excludes the
case of homogeneous magnetic field. We are allowed, however, to consider
magnetic fields which are constant on an arbitrarily large finite domain.
The behavior of the vector potential on the external lines of the graph
away from the vertices does not influence the operator $\Delta(A\cU,B\cU)$
obtained from $\Delta(A,B,\cA)$ by a gauge transformation. Therefore,
constant magnetic fields can be considered as well.

\begin{remark}\label{replace}
We may replace the pair $(A,B)$ by the pair $(CA,CB)$ where $C$ is any invertible $(n+2m)\times
(n+2m)$ matrix. In fact such a replacement does not change the linear conditions \eqref{mag:sa}
and \eqref{2:magnetic} is still satisfied. In other words, the maximal isotropic subspace is left
unchanged and so we have $\Delta(CA,CB;\cA)=\Delta(A,B;\cA)$.
\end{remark}

Let $\chi(x)$ denote an arbitrary family $\{\chi_{j}(x)\}_{j\in\cI\cup\cE}$
of continuously differentiable real valued functions on $[0,a_j]$ if
$j\in\cI$ and on $[0,\infty)$ if $j\in\cE$. To such $\chi$ we associate a
unitary map $G(\chi)$ in $\cH$ by
\begin{equation*}
(G(\chi)\psi)_{j}(x)=\e^{\i\chi_{j}(x)}\psi_{j}(x),\qquad j\in\cI\cup\cE.
\end{equation*}
We claim that
\begin{equation*}
G(\chi)^\dagger \Delta(A,B; \cA)G(\chi)=\Delta(A\cU, B\cU;
\cA-\chi^{\prime})
\end{equation*}
with
\begin{equation*}
\chi^{\prime}_{j}(x)=\frac{d\chi_{j}(x)}{dx}
\end{equation*}
and $\cU$ defined by \eqref{intro:u}. Indeed, setting $\psi = U \phi$ we obtain
\begin{equation*}
A \underline{\psi} + B (\underline{\psi}' + \i
\underline{\cA}\underline{\psi}) = A \cU \underline{\phi} + B \cU
(\underline{\phi}' -\i(\underline{\cA}-\underline{\chi}')\phi) = 0,
\end{equation*}
where $\underline{\chi}'$ is the diagonal matrix given by
\begin{equation*}
\underline{\chi}' = \diag\left(\{\chi_{e}'(0)\}_{e\in\cE},\{\chi_{i}'(0)\}_{i\in\cI},
\{-\chi_{i}'(a_i)\}_{i\in\cI}\right)
\end{equation*}
and $\underline{\cA}$ is defined by \eqref{a:def}. Choosing
$\underline{\chi}'=\cA$ we obtain
\begin{equation*}
G(\chi)^\dagger \Delta(A,B; \cA)G(\chi)=\Delta(A\cU, B\cU).
\end{equation*}

We turn to the proof of Theorem \ref{x:theorem:intro:1}. For the proof we
need the following auxiliary result. Recall that $A^\star$ denotes the
Moore-Penrose pseudoinverse of $A$. Let $P_{\cK}$ be the orthogonal
projection onto the linear subspace $\cK\subset\C^{n+2m}$. Since the matrix
$(A,B)$ has a maximal rank, $P_{\Ran A} + P_{\Ran B}$ is invertible.

\begin{lemma}\label{lemma:2.3}
Assume that $\cU$ leaves both $\Ker A$ and $\Ker B$ invariant. Then there is a matrix $C$
satisfying
\begin{equation}\label{proof:2}
A\cU A^\star = C P_{\Ran A} \qquad\text{and}\qquad B\cU B^\star = C P_{\Ran B},
\end{equation}
if and only if
\begin{equation}\label{compat}
A\cU A^\star y = B\cU B^\star y
\end{equation}
holds for all $y\in \Ran A \cap \Ran B$. If $C$ exists, it is invertible
and given by
\begin{equation}\label{matrix:C}
C = (A\cU A^\star + B\cU B^\star)(P_{\Ran A} + P_{\Ran B})^{-1}
\end{equation}
\end{lemma}

\begin{proof}
Assume there is a matrix $C$ satisfying \eqref{proof:2}. Then
\eqref{compat} and \eqref{matrix:C} are obvious. We claim that $C$ is
invertible. To see this observe that from \eqref{proof:2} and the fact that $\cU$ leaves
$\Ker A$ and $\Ker B$ invariant, it follows that $C$ maps $\Ran A$ and
$\Ran B$ onto themselves. Thus, the range of $C$ is all of $\C^{n+2m}$.

Now assume that \eqref{compat} holds for all $y\in \Ran A \cap \Ran B$. We
prove that the matrix $C$ given by \eqref{matrix:C} satisfies
\eqref{proof:2}. Since $\Ker A^\star = \Ker A^\dagger$ we have $A^\star =
A^\star P_{\Ran A}$. Similarly, $B^\star = B^\star P_{\Ran B}$. Therefore,
\begin{equation}\label{intermit}
C P_{\Ran A} = A \cU A^\star P_{\Ran A} (P_{\Ran A} + P_{\Ran B})^{-1} P_{\Ran A} + B \cU B^\star
P_{\Ran B} (P_{\Ran A} + P_{\Ran B})^{-1} P_{\Ran A}.
\end{equation}

By a result of \cite{Anderson:Schreiber},
\begin{equation*}
P_{\Ran A}(P_{\Ran A} + P_{\Ran B})^{-1} P_{\Ran B} = P_{\Ran B}(P_{\Ran A} + P_{\Ran B})^{-1}
P_{\Ran A} = \frac{1}{2} P_{\Ran A \cap \Ran B}.
\end{equation*}
{}From this we immediately obtain
\begin{equation*}
P_{\Ran A}(P_{\Ran A} + P_{\Ran B})^{-1} P_{\Ran A} = P_{\Ran A} -
\frac{1}{2} P_{\Ran A \cap \Ran B}.
\end{equation*}

Thus, from \eqref{intermit} it follows that
\begin{equation*}
\begin{split}
C P_{\Ran A} & = A \cU A^\star - \frac{1}{2} A \cU A^\star P_{\Ran A \cap
\Ran B} +
\frac{1}{2} B \cU B^\star P_{\Ran A \cap \Ran B}\\
& = A \cU A^\star - \frac{1}{2} A \cU A^\star P_{\Ran A \cap \Ran B} +
\frac{1}{2} A \cU A^\star P_{\Ran A \cap \Ran B} = A \cU A^\star,
\end{split}
\end{equation*}
where we have used \eqref{compat}. The second relation in \eqref{proof:2}
is proved in the same way.
\end{proof}

\renewcommand{\proofname}{Proof of Theorem \ref{x:theorem:intro:1}}

\begin{proof}
By Remark \ref{replace} $\Delta(A\cU,B\cU)=\Delta(A,B)$ if and only if there is an invertible
$(n+2m)\times(n+2m)$ matrix such that
\begin{equation}\label{proof:1}
A\cU = C A\qquad\text{and}\qquad B\cU=CB.
\end{equation}

Assume that these relations hold with some $C$. Then the first relation in
\eqref{proof:1} implies that $A\cU y=0$ for any $y\in\Ker A$. Thus $\cU$
leaves $\Ker A$ invariant. Similarly, by the second relation in
\eqref{proof:1} $\Ker B$ is left invariant by $\cU$.

Since $\Ran A^\star = \Ran A^\dagger$ any $y\in\left[\Ker A\right]^\perp$
can be represented as $y=A^\star z$ with some $z\in\C^{n+2m}$. Therefore,
from \eqref{proof:1} it follows that
\begin{equation}\label{proof:2x}
A\cU A^\star = C P_{\Ran A} \qquad\text{and}\qquad B\cU B^\star = C P_{\Ran
B}.
\end{equation}

Using Lemma \ref{lemma:2.3} we conclude
\begin{equation}\label{compat:x}
A\cU A^\star y = B \cU B^\star y
\end{equation}
for all $y\in \Ran A \cap \Ran B$. Multiplying \eqref{compat:x} by
$A^\star$ from the left and using $A^\star A = P_{\Ran A^\dagger}$ we obtain
\begin{equation}\label{compat:1}
P_{\Ran A^\dagger} \cU A^\star y = A^\star B \cU B^\star y\quad \text{for
any}\quad y\in \Ran A \cap \Ran B.
\end{equation}
Since $\Ran A^\star = \Ran A^\dagger$  and since $\cU$ leaves $\Ran A^\dagger$ invariant, we can
omit the projection on the l.h.s.\ of this equation.

Let $z = B^\star y$. Since $\Ker B^\star=\Ker B^\dagger$ we have $z\neq 0$.
It easy to see that $y=Bz$. Inserting this into \eqref{compat:1} we obtain
\begin{equation*}
\cU A^\star B z = A^\star B \cU P_{\Ran B^\dagger}z = A^\star B \cU z
\end{equation*}
for all $z\in B^\star(\Ran A \cap \Ran B) = B^{-1}(\Ran A \cap \Ran B)$. This proves
\eqref{x:compat:intro:1}. The proof of \eqref{x:compat:intro:2} is similar.

Conversely, let (i) and (ii) be valid. By Lemma \ref{lemma:2.3} the matrix
$C$ defined by \eqref{matrix:C} satisfies \eqref{proof:1}.
\end{proof}

\renewcommand{\proofname}{Proof}

For given boundary conditions $(A,B)$ let $\W(A,B)$ be the set of all $\cU\in\U$ such that
$\Delta(A\cU,B\cU)=\Delta(A,B)$. This set, by construction the stabilizer of the point
$\Delta(A,B)$, is obviously a group by well known arguments from Group
Theory. However, we prefer to give a direct proof.

\begin{proposition}\label{prop:3.4}
$\W(A,B)$ is a group.
\end{proposition}

\begin{proof}
Assume that $\cU_1, \cU_2\in\W=\W(A,B)$. By Theorem \ref{x:theorem:intro:1}
$\cU=\cU_2\cU_1$ leaves both $\Ker A$ and $\Ker B$ invariant. In addition
this theorem implies that
\begin{equation}\label{Zeile}
\cU_1 A^\star B z = A^\star B \cU_1 z,\qquad \cU_2 A^\star B z = A^\star B \cU_2 z
\end{equation}
is valid for all $z\in B^{-1}(\Ran A\cap\Ker B)$. Multiplying the first equation
by $\cU_2$ from the left we obtain
\begin{equation*}
\cU_2\cU_1 A^\star B z = \cU_2 A^\star B \cU_1 z.
\end{equation*}
We claim that $\cU_1 z\in B^{-1}(\Ran A\cap\Ran B)$. Then we will obtain
from the second equation in \eqref{Zeile} the equality $\cU_2\cU_1 A^\star
B z =  A^\star B \cU_2\cU_1 z$ which implies that $\cU_2\cU_1\in\W$.

To prove that $\cU_1 z\in B^{-1}(\Ran A\cap\Ran B)$ it is sufficient to show that $B \cU_1 z \in
\Ran A$. Assume there is $z\in B^{-1}(\Ran A\cap\Ran B)$ such that $B \cU_1 z \perp \Ran A$. Then
from the first equation in \eqref{Zeile} it follows that $A^\star B z = 0$, i.e., $B z \perp \Ran
A$, a contradiction.

The proof that $\cU_1^{-1}\in\W$ is similar and will therefore be omitted.
\end{proof}

{}For local boundary conditions the group $\W$ can be factorized as a direct
product of its subgroups $\W_{\v}$,
\begin{equation}\label{direct:product}
\W(A,B)=\bigtimes_{\v\in V} \W_{\v}(A_{\v},B_{\v}).
\end{equation}

\begin{example}\label{conservative}
An important example of local boundary conditions is given by the matrices
\begin{equation}
\label{ABspecial}
\begin{aligned}
A_{\v}= \begin{pmatrix}
    1&-1&0&\ldots&&0&0\\
    0&1&-1&\ldots&&0 &0\\
    0&0&1&\ldots &&0 &0\\
    \vdots&\vdots&\vdots&&&\vdots&\vdots\\
    0&0&0&\ldots&&1&-1\\
    0&0&0&\ldots&&0&k_{\v}
     \end{pmatrix},\qquad
B_{\v}= \begin{pmatrix}
    0&0&0&\ldots&&0&0\\
    0&0&0&\ldots&&0&0\\
    0&0&0&\ldots&&0&0\\
    \vdots&\vdots&\vdots&&&\vdots&\vdots\\
    0&0&0&\ldots&&0&0\\
    1&1&1&\ldots&&1&1
\end{pmatrix},
\end{aligned}
\end{equation}
where $k_{\v}$ is an arbitrary real number. If $k_{\v}=0$ the matrices
\eqref{ABspecial} define the so-called standard boundary conditions. The
case $k_{\v}\neq 0$ corresponds to the ``delta potential'' of strength
$k_{\v}$ (see, e.g., \cite{Exner}). It is easy to see that the relations
\begin{equation*}
\Ker A_{\v}=\mathrm{linear\ span}\{\begin{pmatrix} k_{\v} \\ k_{\v}\\ k_{\v}\\ \vdots\\ k_{\v} \\
k_{\v}
\end{pmatrix}\},\quad \Ker B_{\v}=\mathrm{linear\ span}\{\begin{pmatrix} 1 \\ -1\\ 0\\ \vdots\\ 0\\ 0
\end{pmatrix}, \begin{pmatrix} 0 \\ 1\\ -1\\ \vdots\\ 0\\ 0
\end{pmatrix},\ldots,\begin{pmatrix} 0 \\ 0\\ 0\\ \vdots\\ 1\\ -1
\end{pmatrix}\}
\end{equation*}
are valid. Moreover,
\begin{equation*}
\Ran A_{\v}\cap\Ran B_{\v} = \Ran B_{\v}=\mathrm{linear\ span}\{\begin{pmatrix} 0
\\ 0\\ 0\\ \vdots\\ 0\\ k_{\v}
\end{pmatrix}\}.
\end{equation*}
Thus, independently of the value of $k_{\v}$, any diagonal unitary
$\cU_{\v}$ leaving both $\Ker A_{\v}$ and $\Ker B_{\v}$ invariant is
necessarily of the form $\cU_{\v}=\e^{\i\phi_{\vsmall}} \1_{\v}$ with
$\1_{\v}$ being the $\deg(\v)\times\deg(\v)$ identity matrix. In either
case the matrices of this form satisfy (ii) of Theorem
\ref{x:theorem:intro:1}. Therefore, $\W_{\v}\cong\SS$ for all $\v\in V$
such that $\W\cong \T^{|V|}$.
\end{example}

\begin{example}\label{Dirichlet}
The case $A=\1$ and $B=0$ provides an example where $\W(A,B)=\U$. Indeed, since $\Ran A\cap\Ran
B=\{0\}$, $\Ker A=\{0\}$, and $\Ker B=\C^{n+2m}$ Theorem \ref{x:theorem:intro:1} implies that any
$\cU\in\U$ belongs to $\W(A,B)$.
\end{example}

We denote the group $\W$ from Example \ref{conservative} by $\W_0$.

\begin{definition}\label{W0}
$\W_0$ is the set of all elements $\cU$ of $\U$ having the property
\begin{equation*}
\cU|_{\cL_{\vsmall}}= \e^{i\phi_{\vsmall}} \1_{\cL_{\vsmall}},\qquad \e^{i\phi_{\vsmall}}\in\SS
\end{equation*}
for all $\v\in V$. Here the $\cL_{\v}$'s are the linear subspaces of $\C^{n+2m}$
defined by \eqref{decomp}.
\end{definition}

We emphasize that the group $W_0$ is independent of the boundary conditions and completely
determined by the graph $\cG$. From Theorem \ref{x:theorem:intro:1} we immediately obtain

\begin{lemma}\label{inclusion}
For arbitrary local self-adjoint boundary conditions $(A,B)$ the group
$\W_0$ is a subgroup of $\W(A,B)$.
\end{lemma}

The groups $\W_0$ and $\U_0$ have a common subgroup consisting of elements $\cU=\e^{\i\varphi}\1$
with $\1$ being the unit $(n+2m)\times(n+2m)$ matrix. Obviously, it is the largest common
subgroup, i.e., $\W_0\cap \U_0 \cong \SS$. Thus, the dimension of $\W_0 \U_0$ is equal to
\begin{equation}\label{dimension:W0U0}
\dim \W_0 \U_0 = \dim \W_0 + \dim \U_0 -1= |V|+|\cI|+|\cE|-1.
\end{equation}

\section{Magnetic Fluxes}\label{sec:3}

In this section we will prove Theorem \ref{intro:thm:2} as well as its Corollary \ref{coro:1}. The
proof uses well-known methods of homological algebra. The relevant concepts used below are
explained in standard textbooks on Algebraic Topology as for example in \cite{Spanier}.

Let $\cG_{\mathrm{int}}$ denote the graph obtained from $\cG$ by removing
its external lines. The graph $\cG_{\mathrm{int}}$ can be viewed as a
one-dimensional simplicial complex.

We consider the (additive) Abelian groups $\cC_1$ and $\cC_0$ generated by the 1-simplices
$\sigma_i$, $i\in\cI$ and the 0-simplices $\sigma_{\v}$, $\v\in V$, respectively, i.e.,
\begin{equation*}
\cC_1 = \left\{ \sum_{i\in\cI} n_i \sigma_i\Big| n_i\in\Z\right\}
\cong\Z^{|\cI|},\qquad \cC_0 = \left\{ \sum_{\v\in V} n_{\v}
\sigma_{\v}\Big| n_{\v}\in\Z\right\} \cong\Z^{|V|}.
\end{equation*}

{}For a given internal line labeled by $i\in\cI$, let $\v_{+}(i)$ be the
terminal vertex which corresponds to the endpoint $a_{i}$ of the interval
$[0,a_{i}]$ and $\v_{-}(i)$ the initial one which corresponds to the other
endpoint $0_{i}$.

We define the boundary operator $\partial_1: \cC_1 \rightarrow \cC_0$
\begin{equation*}
\partial_{1}:\quad c = \sum_{i\in\cI}n_{i} \sigma_{i}\quad
\longmapsto\quad\partial_{1} c=\sum_{i\in\cI}n_{i}
(\sigma_{\v_{+}(i)}-\sigma_{\v_{-}(i)}).
\end{equation*}
Obviously, $\partial_1$ is a group homomorphism. We, obviously, have
\begin{equation}
\label{number} \partial_{1} c=0\quad\Longleftrightarrow\quad
\sum_{i:\;\v_{+}(i)=\v}n_{i}=\sum_{i:\;\v_{-}(i)=\v}n_{i}
\end{equation}
for all $\v\in V(\cG)$. Thus, each element $c\in \Ker\; \partial_{1}$ is a union of closed loops:
$c=\sum_{i\in\cI}n_{i} \sigma_i$ is an oriented closed path, not necessarily connected, where the
internal line $i$ is traversed $|n_{i}|$ times in the positive direction if $n_{i}>0$ and in the
negative direction if $n_{i}<0$.

Also we extend the map $\cC_{1}\xrightarrow{\partial_{1}}\cC_{0}$ to a chain complex
\begin{equation*}
0\longleftarrow\Z\mathop{\longleftarrow}\limits^{\partial_{0}}
\cC_{0}\mathop{\longleftarrow}^{\partial_{1}}\cC_{1}
\end{equation*}
with
\begin{equation*}
\partial_{0}c=\sum_{\v\in V}n_{\v}\quad\mbox{for}\quad c=\sum_{\v\in
V}n_{\v}\sigma_{\v}\in\cC_0.
\end{equation*}
Obviously we have $\partial_{0}\partial_{1}=0$. We call $\Ker \partial_1$
the first homology group $H_1(\cG,\Z)$ of the graph $\cG$ and we set
$H_{0}(\cG,\Z)= \Ker\;\partial_{0}/ \Ran\;\partial_{1}$ to be the zeroth
homology group of $\cG$.

With these preparations we give the definition of the flux map.

\begin{definition}\label{flux:map}
A map $\Phi:\ H_1(\cG,\Z)\rightarrow \SS$ is called a flux map if it is a
group homomorphism, i.e., $\Phi\in\Hom(H_1(\cG,\Z), \SS)$.
\end{definition}

Here $\Hom(H_1(\cG,\Z), \SS)$ denotes the group of all homomorphisms from $H_1(\cG,\Z)$ to $\SS$
with unit element
\begin{equation*}
u:\ c\mapsto 1\in\SS\qquad\text{for\ all}\qquad c\in H_1(\cG,\Z).
\end{equation*}

Observe that $H_1(\cG,\Z)$ is a free Abelian group of finite rank. Its rank
equals the dimension $\dim H_1(\cG,\Z)$ of $H_1(\cG,\Z)$, considered as a
linear space over $\Z$. Since the flux group $\Hom(H_1(\cG,\Z), \SS)$ is
the dual of $H_1(\cG,\Z)$, it is isomorphic to the torus $\T^d$ with
$d=\dim H_1(\cG,\Z)$. To calculate the dimension $d$ we note that the Euler
characteristics $E(\cG_{\mathrm{int}})$ of the simplicial complex
$\cG_{\mathrm{int}}$ equals $|V|-|\cI|$. On the other hand, we have the
standard relation $E(\cG_{\mathrm{int}})=b_0-b_1$, where $b_0=\dim
H_0(\cG,\Z)$ and $b_1=\dim H_1(\cG,\Z)$ are the Betti numbers. The complex
$\cG_{\mathrm{int}}$ is connected since $\cG$ is, so $b_0=1$. Therefore,
\begin{equation}\label{dimension:hom}
\dim \Hom(H_1(\cG,\Z), \SS) = |\cI| - |V| +1.
\end{equation}

Let $\cU$ be an arbitrary diagonal unitary matrix, i.e., $\cU\in\U$. It can be uniquely
represented in the form
\begin{equation}\label{oben}
\cU= \diag\left(\{\e^{\i\varphi_e}\}_{e\in\cE}, \{\e^{\i\varphi_i^-}\}_{i\in\cI},
\{\e^{\i\varphi_i^+}\}_{i\in\cI}\right),
\end{equation}
where the sign ``$-$" corresponds to the initial vertex of the internal edge $i\in\cI$ and the
sign ``$+$" to the terminal vertex.

Define the map $\rho:\ \U \rightarrow \Hom(H_1(\cG,\Z), \SS)$ via
\begin{equation*}
\rho:\ \cU \mapsto \Phi_{\cU}(\cdot)
\end{equation*}
with
\begin{equation}\label{rho:def}
\Phi_{\cU}(c) = \prod_{i\in\cI} \exp\{\i n_i (\varphi_i^+ - \varphi_i^-)\},
\end{equation}
for an arbitrary
\begin{equation*}
c=\sum_{i\in\cI} n_i \sigma_i \in H_1(\cG,\Z).
\end{equation*}
The definition \eqref{rho:def} is obviously consistent with \eqref{intro:flux:prime} and reflects
the additivity of fluxes under the union of loops.

It follows immediately from \eqref{rho:def} that the map $\rho$ is a group
homomorphism. Recall that $\U$ and $\Hom(H_1(\cG,\Z), \SS)$ are isomorphic
to the tori $\T^{n+2m}$ and $\T^d$ with $d=\dim H_1(\cG,\Z)$, respectively.
Any homomorphism from $\T^{n+2m}$ to $\T^d$ can be represented in the form
$\cU \mapsto \exp\{\i M\}\cU$ (component-wise exponentiation) for some
$(n+2m)\times d$ matrix $M$ with integer entries (see, e.g.,
\cite{Broecker}). Thus, the kernel of the homomorphism $\rho$ is a subgroup
isomorphic to a torus.

\begin{lemma}\label{epimorphism}
$\rho$ is an epimorphism.
\end{lemma}

\begin{proof}
We have to prove that the map $\rho$ is surjective. Let $c_p$, $1\leq p
\leq d$ with $d=\dim H_1(\cG,\Z)=\dim \Hom(H_1(\cG,\Z), \SS)$ be an
arbitrary basis of generators.  It suffices to show that for arbitrarily
prescribed $(\e^{\i \mu_1},\ldots,\e^{\i \mu_d})\in\T^d$ there is a
$\cU\in\U$ such that $\rho(\cU)=\Phi_{\cU}$ with $\Phi_{\cU}$ given by
\eqref{rho:def} which satisfies
\begin{equation}\label{blanc}
\Phi_{\cU}(c_p) =\e^{\i \mu_p},\qquad 1\leq p \leq d.
\end{equation}
All generators $c_p$ are of the form $c_p=\sum_{i\in\cI} n_i(p) \sigma_i$
with $n_i(p)\in\Z$.

Let $H_1(\cG,\R)$ be the linear space over $\R$ generated by the basis
$\{c_p\}_{p=1}^d$. We set
\begin{equation*}
\underline{\varphi} =(\varphi_1^-,
\ldots,\varphi_m^-,\varphi_1^+,\ldots,\varphi_m^+)^T \in \R^{2m}.
\end{equation*}
Consider the linear map $S:\ \R^{2m}\rightarrow H_1(\cG,\R)$ given as
\begin{equation*}
S \underline{\varphi} = \sum_{p=1}^d \sum_{i\in\cI} (\varphi_i^+
-\varphi_i^-) n_i(p) c_p.
\end{equation*}
Obviously, if the equation
\begin{equation*}
S \underline{\varphi} = \sum_{p=1}^d \mu_p c_p
\end{equation*}
has a solution, then any matrix $\cU$ defined by $\underline{\varphi}$
through equation \eqref{oben} satisfies \eqref{blanc}.
To prove that this equation indeed has a solution for arbitrary choice of
numbers $\mu_p$ by the Fredholm Alternative it suffices to show that
$\Ker\,S^\dagger=\{0\}$. Assume that
\begin{equation*}
c=\sum_{p=1}^d \alpha_p\, c_p\in \Ker\, S^\dagger,\qquad \alpha_p\in\R.
\end{equation*}
Then
\begin{equation*}
\sum_{p=1}^d \alpha_p\, n_i(p) = 0\qquad \text{for all}\qquad i\in\cI.
\end{equation*}
Hence, we obtain
\begin{equation*}
\sum_{i\in\cI} \Big(\sum_{p=1}^d \alpha_p n_i(p) \Big) \sigma_i =
\sum_{p=1}^d \alpha_p c_p = 0,
\end{equation*}
which implies that $c=0$.
\end{proof}

Using the First Homomorphism Theorem from Lemma \ref{epimorphism} it follows
that
\begin{equation}\label{iso}
\Hom(H_1(\cG,\Z), \SS) \cong \U/\Ker\, \rho.
\end{equation}

Obviously, $\U_0\subseteq \Ker\,\rho$. Also one can easily prove that $\W_0\subseteq \Ker\,\rho$.
Actually, we have the stronger result:

\begin{lemma}\label{thm:4.3}
$\Ker\,\rho = \U_0 \W_0$.
\end{lemma}

\begin{proof}
First we prove that $\W_0\subseteq \Ker\,\rho$. Let $c_1,\ldots,c_d$ be any basis in $H_1(\cG,\Z)$.
It suffices to show that $\rho(\cU)(c_p)=1$ for any $\cU\in\W_0$ and any basis element $c_p$. As
discussed above, $c_p$ can be viewed as a closed loop of length $L$, i.e. a sequence of triples
$(\v(\ell),i(\ell),\v'(\ell))$, $\ell=1,\ldots,L$ with $\v(1)=\v'(L)$. Let
$\cU=\diag(\e^{\i\varphi_1},\ldots,\e^{\i\varphi_{n+2m}})\in\W_0$ with $n=|\cE|$ and $m=|\cI|$.
Since $\cU\in\W_0$ all phases $\varphi_j$ can be labelled by the set of vertices $V$ of the graph
$\cG$. Obviously,
\begin{equation*}
\rho(\cU)(c_k) = \exp\left\{\i\sum_{\ell=1}^L
(\varphi_{\v'(\ell)}-\varphi_{\v(\ell)}) \right\}=
\exp\left\{\i(\varphi_{\v'(L)}-\varphi_{\v(1)})\right\} = 1.
\end{equation*}
Thus, we have proved that $\W_0$ is a subgroup of $\Ker\,\rho$. The
inclusion $\U_0\subseteq \Ker\, \rho$
is obvious. Therefore, $\U_0 \W_0\subseteq \Ker\,\rho$.

Now we calculate the dimension of $\Ker\,\rho$. From \eqref{iso} using \eqref{dimension:hom} it
follows that
\begin{equation*}
\begin{split}
\dim\Ker\, \rho &= \dim \U - \dim \Hom(H_1(\cG,\Z),\SS)\\
&= |\cE| + 2|\cI| - |\cI| + |V| -1\\
&= |\cE| + |\cI| + |V| -1.
\end{split}
\end{equation*}
Recall that by \eqref{dimension:W0U0} $\dim \W_0 \U_0= |\cE| + |\cI| + |V|
-1$. Therefore, the groups $\U_0 \W_0$ and $\Ker\,\rho$ have the same
dimensions. Since both groups are isomorphic to tori and $\U_0 \W_0\subseteq
\Ker\,\rho$, we have $\U_0 \W_0= \Ker\,\rho$.
\end{proof}

Actually, at this stage the proof of the second claim of Theorem \ref{intro:thm:2} is already
completed. Indeed, this follows from the relation \eqref{iso} and Lemma \ref{thm:4.3}.

We turn to the case when $\W=\W(A,B)$ is strictly larger than
$\W_0$. Whereas $W_0\subseteq
\W(A,B)$ for all local boundary conditions, for non-local boundary conditions it may well happen
that $\W(A,B)\subsetneq \W_0$ (see Example \ref{ex:4.1} below).

Let $\theta$ be the canonical quotient map $\Hom(H_1(\cG;\Z),\SS)\rightarrow
\Hom(H_1(\cG;\Z),\SS)/\rho(\W)$ with $\Ker\,\theta = \rho(W)$. Consider the
composition
\begin{equation*}
\widehat{\rho}=\theta \circ \rho.
\end{equation*}

\renewcommand{\proofname}{Proof of Theorem \ref{intro:thm:2}}

\begin{proof}
By construction $\widehat{\rho}$ is a homomorphism. Recall that
$\W_0\subseteq\W(A,B)$ for local boundary conditions $(A,B)$. Lemma
\ref{thm:4.3} implies that $\Ker\,\widehat{\rho}=\U_0 \W(A,B)$. {}From
Lemma \ref{epimorphism} it follows that $\Ran\,\widehat{\rho} =
\Hom(H_1(\cG;\Z),\SS)/\rho(\W)$. Applying the First Homomorphism Theorem we
obtain the claim.
\end{proof}

\renewcommand{\proofname}{Proof of Corollary \ref{coro:1}}

\begin{proof}
Since $\cG_{\mathrm{int}}$ is a tree, the group $H_1(\cG,\Z)$ is trivial.
Thus, the flux group $\Hom(H_1(\cG,\Z), \SS)$ is also trivial. Now Theorem
\ref{intro:thm:2} implies that $\U=\U_0 \W(A,B)$.
\end{proof}

The requirement of locality of boundary conditions in the statement of Corollary \ref{coro:1} is
crucial since non-local boundary conditions may act like a loop. This can be illustrated by the
following example.

\begin{figure}[ht]
\centerline{ \unitlength1mm
\begin{picture}(120,40)
\put(20,20){\line(1,0){80}} \put(50,20){\circle*{2}} \put(70,20){\circle*{2}} \put(40,21){1}
\put(60,21){2} \put(80,21){3} \put(60,20){\vector(1,0){2}}
\end{picture}}
\caption{The graph from Example {\protect\ref{ex:4.1}}. It has two vertices, one internal, and
two external lines. The internal line has length $a$, the arrow shows its
orientation.}\label{fig:line}
\end{figure}
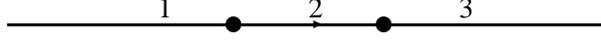

\begin{example}\label{ex:4.1}
Consider the graph depicted in Fig.\ \ref{fig:line}. Consider the magnetic Laplacian
$\Delta_{\cA}$ with the boundary conditions
\begin{equation*}
A\underline{\psi}+B (\underline{\psi}'-\i\underline{\cA}\underline{\psi})=0,
\end{equation*}
where
\begin{equation}\label{AB}
A=\begin{pmatrix} 1 & -1 & 0 & 0 \\ 0 & 1 & -1 & 0 \\ 0 & 0 & 1 & 0 \\ 1 & 0 & 0 & 0
\end{pmatrix},\qquad B=\begin{pmatrix} 0 & 0 & 0 & 0 \\ 0 & 0 & 0 & 0 \\
1 & 1 & 0 & 0 \\ 0 & 0 & 1 & 1
\end{pmatrix}.
\end{equation}
Here we use the following ordering: $\underline{\psi}=(\psi_1(0), \psi_2(0),
\psi_2(a),\psi_3(0))^T$. The boundary conditions \eqref{AB} are obviously non-local in the sense of
Definition \ref{def:local}.

It is easy to check that
\begin{equation*}
\begin{split}
\Ker A=\{0\},\qquad &\Ker B=\mathrm{linear\ span}\{
\begin{pmatrix} 0 \\ 0 \\ 1 \\ -1 \end{pmatrix},
\begin{pmatrix} 1 \\ -1 \\ 0 \\ 0 \end{pmatrix}\},\\
\Ran A=\C^4,\qquad &\Ran B=\mathrm{linear\ span}\{
\begin{pmatrix} 0 \\ 0 \\ 1 \\ 0 \end{pmatrix},
\begin{pmatrix} 0 \\ 0 \\ 0 \\ 1 \end{pmatrix}\}.
\end{split}
\end{equation*}

Assume there is a gauge transformation $G$ such that the pairs $(A\cU, B\cU)$ and $(A,B)$ with
$\cU=\cU_G\in \U$ define the same Laplacian. By Theorem \ref{x:theorem:intro:1} since $\cU$ leaves
$\Ker A$ and $\Ker B$ invariant, it has the form
\begin{equation*}
\cU=\diag (\e^{\i\phi_1}, \e^{\i\phi_1}, \e^{\i\phi_2}, \e^{\i\phi_2}).
\end{equation*}
We have
\begin{equation*}
A^\star B = \begin{pmatrix} 0 & 0 & 1 & 1 \\ 0 & 0 & 1 & 1\\ 1 & 1 & 0 & 0\\ 1 & 1 & 0 & 0
\end{pmatrix}.
\end{equation*}
Simple calculations yield $[\cU,A^\star B]z=0$ for all
\begin{equation*}
z\in B^{-1}\Ran B = \mathrm{linear\ span}\{
\begin{pmatrix} 1 \\ 1 \\ 0 \\ 0 \end{pmatrix},
\begin{pmatrix} 0 \\ 0 \\ 1 \\ 1 \end{pmatrix}\}
\end{equation*}
if and only if $\phi_1=\phi_2$ modulo $2\pi$. Therefore the boundary conditions $(A\cU, B\cU)$ and
$(A,B)$ are equivalent if and only if
\begin{equation*}
\int_0^a \cA(x) dx = 0\quad\text{modulo}\quad 2\pi.
\end{equation*}

Note that the isotropy group $\W(A,B)$ consists of all $4\times 4$ unitary diagonal matrices which
are a multiple of the unit matrix. Obviously, $\W(A,B)$ is a proper subgroup of $\W_0$.
\end{example}


\end{document}